\documentclass[a4paper,fleqn,usenatbib]{mnras}

\usepackage[T1]{fontenc}
\usepackage{ae,aecompl}

\usepackage{graphicx}	
\usepackage{amsmath}	
\usepackage{amssymb}	
\usepackage{geometry,graphicx,kantlipsum,rotating}   
\usepackage{tikz}
\usepackage{adjustbox}

\title[\textbf{Timing stability of black widow pulsars}]{Timing stability of three black widow pulsars}

\author[A. Bak Nielsen et al.]{
Ann-Sofie Bak Nielsen$^{1,2,4}$\thanks{E-mail: nielsen@strw.leidenuniv.nl},
Gemma H. Janssen$^{3,5}$,
Golam Shaifullah$^{3}$,\newauthor
Joris P. W. Verbiest$^{2,1}$,
David J. Champion$^{1}$,
Gr\'egory Desvignes$^{11}$,
Lucas Guillemot$^{6,7}$,
\newauthor 
Ramesh Karuppusamy$^{1}$,
Michael Kramer$^{1}$,
Andrew G. Lyne$^{8}$,
Andrea Possenti$^{9,12}$,
\newauthor
Ben W. Stappers$^{8}$,
Cees Bassa$^{3}$,
Isma\"el Cognard$^{6,7}$,
Kuo Liu$^{1}$,
Gilles Theureau$^{6,7,10}$
\\
$^{1}$ Max-Planck-Institut f{\"u}r Radioastronomie, Auf dem H{\"u}gel 69, 53121 Bonn, Germany\\
$^{2}$ Fakult{\"a}t f{\"u}r Physik, Universit{\"a}t Bielefeld, Postfach 100131, 33501 Bielefeld, Germany\\
$^{3}$ ASTRON, the Netherlands Institute for Radio Astronomy, Oude Hoogeveensedijk 4, 7991 PD Dwingeloo, The Netherlands\\
$^{4}$ Leiden Observatory, Leiden University, Niels Bohrweg 2, 2333 CA, Leiden\\ 
$^{5}$ Department of Astrophysics/IMAPP, Radboud University, P.O. Box 9010, 6500 GL Nijmegen, The Netherlands\\ 
$^{6}$ Laboratoire de Physique et Chimie de l'Environnement et de l'Espace LPC2E CNRS-Universit{\'e} d'Orl{\'e}ans, 45071 Orl{\'e}ans, France\\
$^{7}$ Station de radioastronomie de Nan{\c c}ay, Observatoire de Paris, PSL Research University, CNRS/INSU 18330 Nan{\c c}ay, France\\
$^{8}$ Jodrell Bank Centre for Astrophysics, School of Physics and Astronomy, The University of Manchester, Manchester M13 9PL, UK\\
$^{9}$ INAF - Osservatorio Astronomico di Cagliari, via della Scienza 5, 09047 Selargius (CA), Italy \\
$^{10}$ LUTH, Observatoire de Paris, PSL Research University, CNRS, Universit\'e Paris Diderot, Sorbonne Paris Cit\'e, 92195 Meudon, France\\
$^{11}$LESIA, Observatoire de Paris, Universit\'e PSL, CNRS, Sorbonne Universit\'e, Universit\'e de Paris,\newline 5 place Jules Janssen, 92195 Meudon, France\\
$^{12}$ Universit\`a di Cagliari, Dept Physics, S.P. Monserrato-Sestu Km 0,700 - 09042 Monserrato (CA), Italy}
\date{Accepted XXX. Received YYY; in original form ZZZ}

\pubyear{2020}

\begin{document}
\label{firstpage}
\pagerange{\pageref{firstpage}--\pageref{lastpage}}
\maketitle

\begin{abstract}
We study the timing stability of three black widow pulsars, both in terms of their long-term spin evolution and their shorter-term orbital stability. The erratic timing behaviour and radio eclipses of the first two black widow pulsar systems discovered (PSRs B1957+20 and J2051$-$0827) was assumed to be representative for this class of pulsars. With several new black widow systems added to this population in the last decade, there are now several systems known that do not show these typical orbital variations or radio eclipses.
We present timing solutions using 7$-$8 yrs of observations from four of the European Pulsar Timing Array telescopes for PSRs\,J0023+0923, J2214+3000 and J2234+0944,
and confirm that two of these systems do not show any significant orbital variability over our observing time span, both in terms of secular or orbital parameters. The third pulsar PSR J0023+0923 shows orbital variability and we discuss the implications for the timing solution.
Our results from the long-term timing of these pulsars provide several new or improved parameters compared to earlier works. We discuss our results regarding the stability of these pulsars, and the stability of the class of black widow pulsars in general, in the context of the binary parameters, and discuss the potential of the Roche-lobe filling factor of the companion star being an indicator for stability of these systems.
\end{abstract}


\section{Introduction}

The first radio pulsar was discovered in 1967 by Jocelyn Bell Burnell \citep{Hewish_1968Natur.217..709H} and since then more than 2700 radio pulsars have been found \citep{Manchester_2005yCat.7245}. This sample includes both isolated pulsars and pulsars in binary systems, and slow and millisecond pulsars (MSPs). 
MSPs are old pulsars that have gone through the recycling process \citep{Alpar_1982Natur.300..728A, Bhattacharya_1991PhR...203....1B}. 
MSPs are spun up in X-ray binary systems through accretion, which transfers angular momentum onto the pulsar. During the X-ray pulsar phase, the magnetic field strength of the pulsar decreases \citep{Chen_2013ApJ...775...27C}.
A special group of MSPs, the so-called `spiders', with very low companion masses and tight orbits (P$_b$<24\thinspace hr), are the `redback' pulsars (RBPs), which have companion star masses of M$_{2}\simeq 0.1-0.4\thinspace $M$_{\odot}$, and the `black widow' pulsars (BWPs) with M$_{2}\ll 0.1\thinspace $M$_{\odot}$ \citep{Chen_2013ApJ...775...27C,Roberts_2013IAUS..291..127R}.
The low mass companions of BWPs are often thought to be semi-degenerate stars, similar to brown dwarfs (see section 4.4 in \citealt{Lazaridis_2011MNRAS.414.3134L}). The companions of RBPs are probably non-degenerate stars, as suggested by a few cases where the optical counterpart was discovered \citep{Ferraro_2001ApJ...561L..93F, Roberts_2013IAUS..291..127R, AlNoori_2018ApJ...861...89A}. The fate of the companion star in the BWP systems is unclear. However, one postulated scenario is that the companion could be fully ablated, creating an isolated MSP. Alternatively,
tidal effects could play a significant role in eventually destroying the companion \citep{Stappers_1998ApJ...499L.183S, Chen_2013ApJ...775...27C}.

Since the discoveries of PSRs B1957+20 and J2051$-$0827 \citep{Fruchter_1988Natur.333..237F, Stappers_1996ApJ...473L.119S}, the populations of BWPs and RBPs have increased greatly due to recent surveys, using either radio observations which were optimised to find MSPs, or targeted radio surveys which are guided by high energy point sources such as those seen with the gamma-ray observatory \textit{Fermi}. More than 40 MSPs were found using such surveys \citep{Roberts_2013IAUS..291..127R}. The two original BWP systems show eclipses of the radio emission from the pulsar. The eclipses are due to material being ablated off of the companion star by a strong pulsar wind \citep{Fruchter_1988Natur.333..237F, Stappers_1996ApJ...473L.119S, Chen_2013ApJ...775...27C}. The presence of this matter in the binary, often makes the BWPs complicated systems to use for high precision pulsar timing experiments. The companion stars are often bloated, and their unstable and in-homogeneous mass distributions exerts small torques on the pulsar, which in turn causes small changes in the orbit. Such orbital variations increase the number of parameters that are required in timing models, and mean that it is not possible to predict the timing behaviour and thus not possible to model the orbital variations for longer timescales. The orbital variability could also reduce the sensitivity of the timing solution to other signals of interest, such as for example gravitational-wave (GW) signals.

By observing multiple stable pulsars in different parts of the sky, an array of pulsars, 
and cross-correlating their timing residuals, it should be possible to obtain the sensitivity required to measure the expected quadrupolar signature of a nHz GW. This is the general idea behind the pulsar timing arrays (PTAs) \citep{Foster_1990BAAS...22.1341F, Tiburzi_2018PASA...35...13T}. With new BWPs being discovered, \citet{Bochenek_2015ApJ...813L...4B} considered the feasibility of using BWPs in PTAs, by testing if fitting for multiple orbital frequency derivatives significantly reduced the sensitivity to the GW signal. 
They used simulated data sets of five different pulsars to constrain how sensitive a timing solution would be to a GW signal, if several orbital-period derivatives were fitted. 
They found that the sensitivity to the GW signals was not reduced significantly, partly due to the comparatively long periods of the GW signals and the typical BWP orbital period of 2 to 20\thinspace hr. 
Recently a few BWPs have been included in PTAs, e.g. PSR J0610$-$2100 is already used by the EPTA \citep{Desvignes_2016MNRAS.458.3341D}. 
There are three other BWPs that show promising timing precision and which are candidates for use in a PTA: PSRs J0023+0923, J2214+3000, and J2234+0944. All three are currently observed by the North American Nanohertz Observatory for Gravitational Waves (NANOGrav, \citealt{Arzoumanian_2018ApJS..235...37A}) and the European Pulsar Timing Array (EPTA, this work).\\

PSR J0023+0923 was discovered with the Green Bank Telescope (GBT) in a survey of unassociated \textit{Fermi} $\gamma$-ray sources \citep{Hessels_2011AIPC.1357...40H}. It is an MSP with a spin period of $\sim$3\thinspace ms, an orbital period of about 3.3\thinspace hr and a minimum companion mass of 0.017\thinspace M$_{\odot}$. 
The optical counterpart to PSR J0023$+$0923 was discovered by \citet{Breton_2013ApJ...769..108B} with the \textit{Gemini North} telescope. By combining measurements of the filling factor with the estimated distance from the dispersion measure value they inferred that the companion star never fills its Roche lobe. Their results show that, when taking the small filling factor into account, the companion star could be as small as 0.05\thinspace $R_{\odot}$. PSR J0023+0923 has also been observed in X-rays, but no pulsations were detected \citep{Ransom_2011ApJ...727L..16R}.\\

Both PSRs J2214+3000 and J2234+0944 were found in radio searches in the directions of unassociated \textit{Fermi}-LAT sources, with the GBT and the Parkes radio telescope respectively \citep{Ransom_2011ApJ...727L..16R, Ray_2012arXiv1205.3089R}. PSR J2214+3000 was discovered to be a BWP system, with a low-mass companion ($\sim$0.014\thinspace $M_\odot$) and an orbital period of about 9.9\thinspace hr and a spin period of 3.12\thinspace ms. The companion star of PSR J2214+3000 was detected in the optical by \citet{Schroeder_2014ApJ...793...78S}. PSR J2234+0944 has an orbital period of about 10.1\thinspace hr, with a companion star of $\sim$0.015\thinspace M$_{\odot}$, and a spin period of $\sim$3.36\thinspace ms \citep{Ray_2012arXiv1205.3089R}. 
Phase resolved $\gamma$-ray pulsations have been detected for all three pulsars \footnote{\texttt{https://confluence.slac.stanford.edu/display/GLAMCOG/\\Public+List+of+LAT-Detected+Gamma-Ray+Pulsars}}\citep{Abdo_2013ApJS..208...17A}.\\

In this paper we present updated timing solutions for the three BWPs J0023+0923, J2214+3000, and J2234+0944. In Section \ref{sec:observations} we present the observations of the three BWPs, in Section \ref{sec:results} we present our timing solutions and in Section \ref{sec:discussion} we discuss the stability of these pulsars and compare their properties with other BWPs. The data used in this paper is available online \footnote{\texttt{http://www.epta.eu.org/}}.

\section{Observations}\label{sec:observations}

\begin{figure}
\centering
\includegraphics[width=0.9\linewidth]{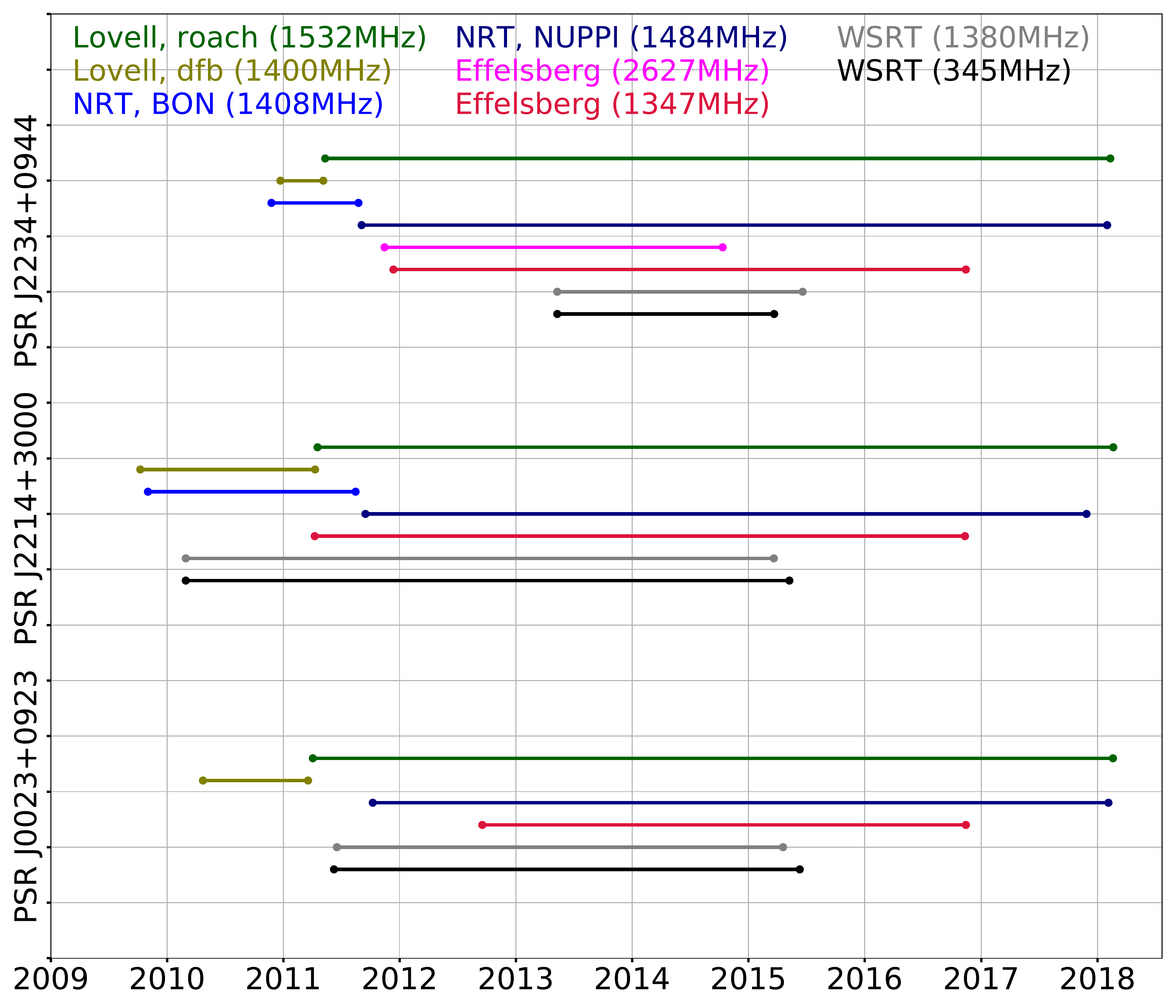}
\caption{Timing baselines per backend for the three BWPs discussed in this paper. The colours correspond to backend and observing frequency combinations. The dark green lines represent the time span covered by the ROACH-based recorder, observing at a centre frequency of 1532$\thinspace$MHz, and the brown line represents the DFB at 1400$\thinspace$MHz, both at the Lovell radio telescope. The light blue line represents the BON backend at 1408$\thinspace$MHz, and the dark blue the NUPPI backend at 1404$\thinspace$MHz, both at the Nan\c{c}ay Radio Observatory. The pink and red lines represent timing spans of the PSRIX backend at 1347 and 2627$\thinspace$MHz, respectively, at the Effelsberg 100-m radio telescope. Finally, the grey and black lines represent timing baselines for the PuMa-II backend at the Westerbork Synthesis Radio Telescope. More details about the backends and telescopes can be found in section \ref{sec:observations} and \autoref{tab:telescopes}.}
\label{fig:time_backed}
\end{figure}

The data in this paper consists of times of arrival (TOAs), collected and observed in the context of the European Pulsar Timing Array (EPTA). The EPTA is a collaboration of scientists, the observational part of which is supported by the data collected by five different telescopes around Europe: the Westerbork Synthesis Radio Telescope (WSRT) in the Netherlands, the Effelsberg 100-meter Radio Telescope (EFF) in Germany, the Lovell Radio Telescope (JBO) in the UK, the Nan\c cay Radio Telescope (NRT) in France, and the Sardinia Radio Telescope (SRT) in Italy \citep{Desvignes_2016MNRAS.458.3341D}. As the SRT has only just started observing on a regular basis, data from this telescope are not used in this paper. 
The three data sets for PSR J0023+0923, J2214+3000 and J2234+0944 span about 7.5, 8 and 7 years of data, respectively, see Fig. \ref{fig:time_backed}. As described in Table \ref{tab:telescopes} the data used are from several pulsar recording instruments \citep{Karuppusamy_2008PASP..120..191K,Desvignes_2011AIPC.1357..349D, Lazarus_2016MNRAS.458..868L,Guillemot_2016A&A...587A.109G,Bassa_2016MNRAS.456.2196B}.\\

The data were collected using dual-polarization receivers, and backends providing the capabilty of coherent dedispersion (except for data from the Lovell digital filterbank (DFB) backend, which employs incoherent dedispersion), thus recording four coherently dedispersed timeseries (one for each Stokes parameter), integrated modulo the spin period (i.e., `online folding`) into subintegrations of between 10-60 seconds, depending on the telescope and date of observation (see Table \ref{tab:telescopes}).

Specific parts of the data that were affected by radio frequency interference (RFI) and not detected or cleaned in the preprocessing stages were cleared using standard \texttt{PSRCHIVE} tools \citep{vanStraten_2012AR&T....9..237V}, where it should be noted, that the Lovell ROACH backend uses a real-time spectral kurtosis \citep{Nita_2010MNRAS.406L..60N} RFI removal method as implemented in {\tt dspsr} in the pre-processing of the data. The cleaned data were then averaged in frequency and time to maximise the signal-to-noise ratio (S/N).
For PSR\,J0023$+$0923, some observations were taken with a total observing length of several hours, to cover the full orbital period. We split observations that were more than 1\thinspace hr long into smaller subsets of 30\thinspace min each, to avoid smearing as a result of a non-optimal timing solution. Due to its short orbital period, data for PSR J0023+0923 were averaged in time several times in an iterative process where in each iteration the timing model was improved. This was needed to optimally adjust the orbital delays when time-averaging the data. \\

Standard templates were constructed for each pulsar by fitting von Mises functions to the total-intensity profiles formed by adding a large number of the averaged observations from the previous step. The TOA for each averaged observation was measured by cross correlation with the respective standard template in the Fourier domain \citep{Taylor_1992RSPTA.341..117T} and uncertainties of the TOAs were estimated using the Fourier domain Markov chain (FDM) Monte Carlo algorithm of the \texttt{pat} tool \citep{vanStraten_2012AR&T....9..237V}.

\begin{figure*}
\includegraphics[width=1.0\linewidth]{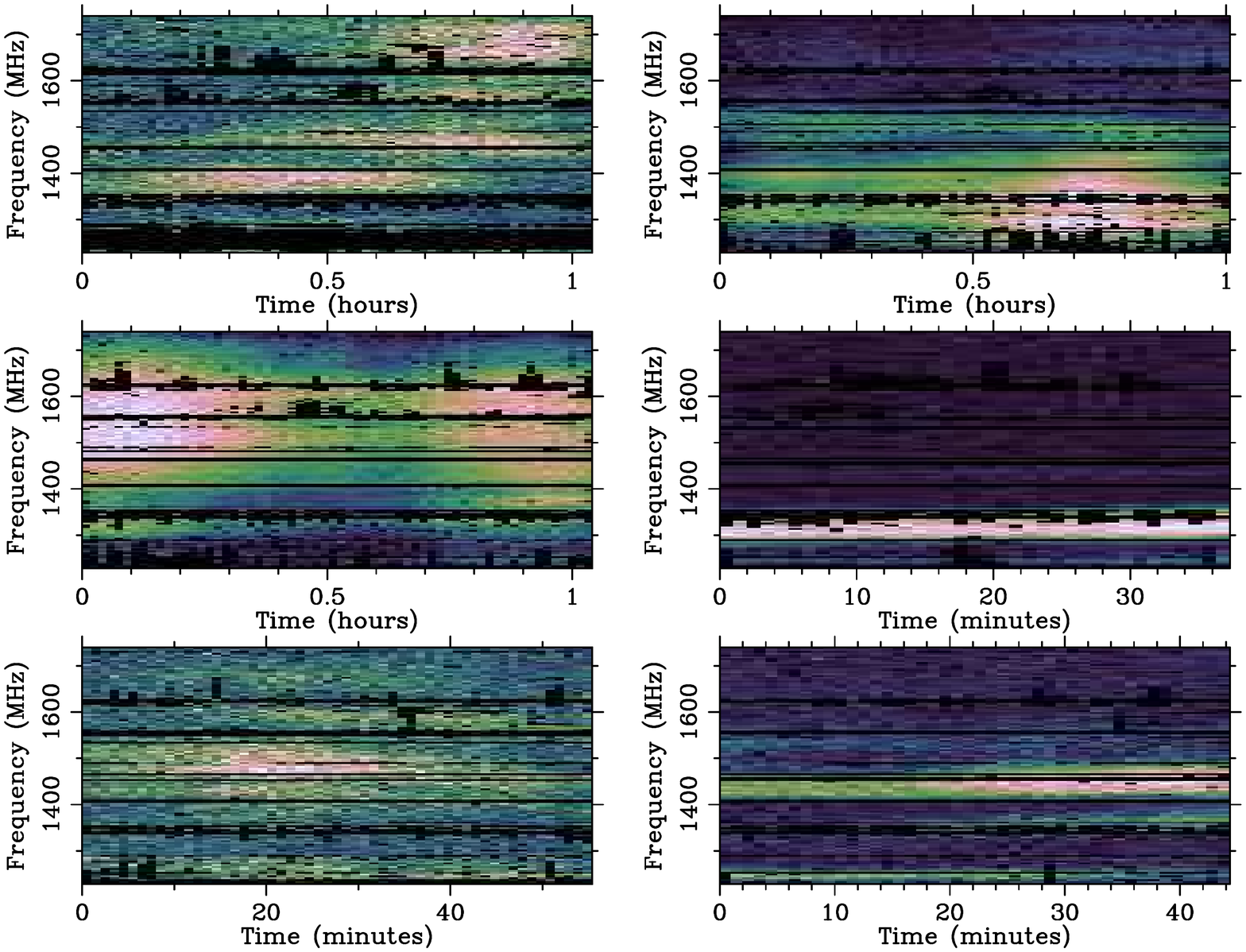}
\caption{Dynamic spectra obtained with Nan\c cay and the NUPPI backend; the scintillation in time and frequency for PSR J2234+0944 is clearly seen, and it affects the timing because the profile shape changes with frequency. The MJDs represented in this plot are 55879, 55888, 55902, 55914, 56046 and 56060. The dark area corresponds to RFI masked data and the bright areas correspond to high intensity scintels. 
}
\label{fig:scintillation}
\end{figure*}

Since the interstellar medium (ISM) is not only magnetised and ionised but also turbulent and inhomogeneous, this modulates the observed flux density as a function of time and observing frequency \citep{Rickett_1969Natur.221..158R}. This effect, known as interstellar scintillation, leads to the pulsar emission being observed across the wide bandwidths of the receivers employed at varying intensities. This is particularly significant for the NUPPI backend at NRT where data is recorded with 512\thinspace MHz of bandwidth around a centre frequency of 1484\thinspace MHz, leading to the `patchy' dynamic spectra shown in Fig. \ref{fig:scintillation}. Integration along the frequency axis leads to TOAs which are referred to the centre of the band rather than the frequency at which the maximum intensity occurs. Standard profiles which do not account for this frequency-dependent feature therefore lead to biased estimates of the TOAs. To account for scintillation, the NUPPI data for PSRs~J0023+0923 and J2214+3000 were split in two bands and into four bands for PSR~J2234+0944.

\begin{table*}
	\centering
	\caption{Telescope (backends), centre frequency, bandwidth, number of TOAs for the different pulsars, range of the error-scaling factor (T2EFAC) and the subintegration length. The four telescopes used are part of the EPTA and they are the Westerbork telescope (WSRT), Effelsberg, Nan\c cay, and Lovell at Jodrell Bank. $^*$The data from 
    Nan\c cay, NUPPI, are affected by scintillation, and the frequency bands have thus been split into smaller bands, of 256\thinspace MHz for PSR J0023+0923 and J2214+3000. For PSR J2234+0944 we have split the band in four, so we have a bandwidth of 128\thinspace MHz. The given T2EFAC range is for the range used for all three pulsars. Furthermore, the sub-integration length is given in the last coloumn.}
\label{tab:telescopes}
	\begin{tabular}{lccccccr} 
		\hline\hline
		Telescope  & f$_c$  & BW &  & No. of TOAs  &  & T2EFAC range & Sub.\thinspace int. time \\
        (Backend) & (MHz) & (MHz) &  J0023+0923 & J2214+3000  & J2234+0944 & & (sec) \\ 
		\hline
		WSRT (PuMaII) & 345 & 70 & 58  & 51 & 23 & 1.1$-$2.4 & 60 \\
        	 & 1380 & 160 & 30  & 60 & 43 & 1.0$-$1.6 & 60 \\
		Effelsberg (PSRIX) & 1347.5 & 200 & 57  & 23 & 32 & 1.6$-$4.5 & 10 \\
        				  & 2627 & 200 & -  & - & 29 & $\sim$1.2 & 10 \\
       	Nan\c cay (NUPPI) & 1484 & 512/256/128$^*$ & 791  & 184 & 635 & 1.4$-$2.1 & 15$-$60\\
        Nan\c cay (BON) & 1408 & 128 & - & 108  & 31 & 1.2$-$1.4 & 30$-$120\\
        Lovell (DFB) & 1400 & 384-512 & 23  & 96 & 47 & 1.2$-$1.6 & 10 \\
        Lovell (ROACH) & 1532 & 400 & 114  & 126 & 146 & 1.1$-$1.8  & 10 \\
		\hline
		\hline
	\end{tabular}
\end{table*}

The TOAs were measured at each observatory with reference to a local clock, which was synchronised using GPS to the Terrestrial Time  standard, that is derived from the ``Temps Atomique International" time standard, TT(TAI). The TOAs were corrected to the Solar-System barycentre using the NASA-JPL DE421 planetary ephemerides \citep{Folkner_2009IPNPR.178C...1F}. The \texttt{Tempo2} software package was used \citep{Hobbs_2006MNRAS.369..655H} to fit the timing solutions presented in Table \ref{tab:parameters}. Data sets generated by different telescopes were phase-aligned by using constant offsets (JUMPS; following the procedure described in \citealt{Verbiest_2016MNRAS.458.1267V}) and a few clock offsets were applied where necessary (see also \citealt{Desvignes_2016MNRAS.458.3341D} and \citealt{Lazarus_2016MNRAS.458..868L})\footnote{Throughout this work \texttt{Tempo2} version 2014.2.1 was used.}.

Error-scaling factors (called \texttt{T2EFACs}) were calculated for each receiver-frequency-band combination using the timing model derived and taking the square root of the reduced $\chi^2$ from fits to the TOAs for each such combination alone. The \texttt{T2EFAC} value corrects the statistical weights of the data by re-scaling the error bars to account for probable under estimation at the template matching stage.

\section{Results}\label{sec:results}

\begin{figure*}
    \centering
    \centering
    \includegraphics[width=1.0\linewidth]{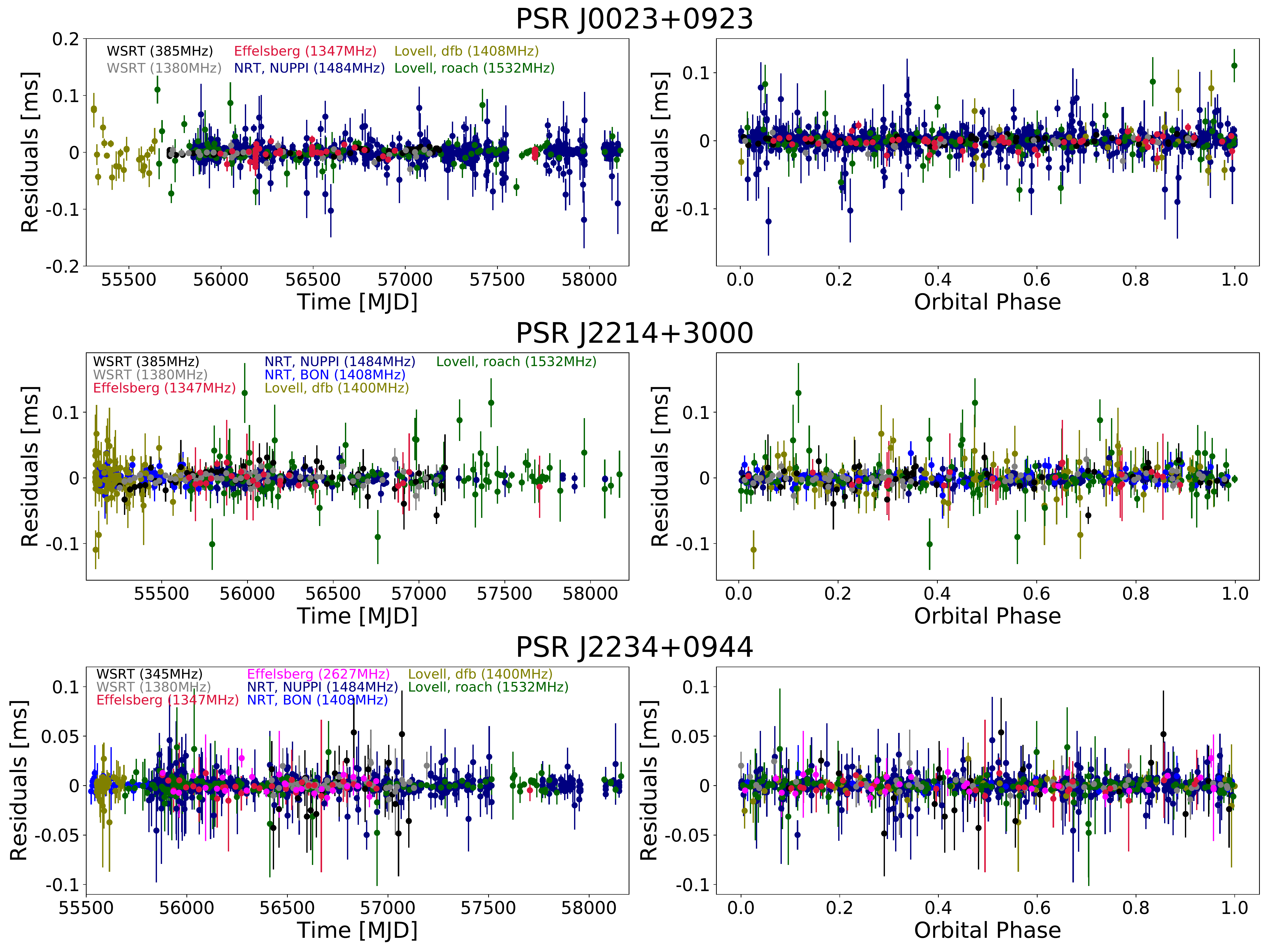}
    \caption{\textbf{Left panels}: The timing residuals of PSRs~J0023+0923, J2234+0944 and J2214+3000 after fitting the timing models presented in Table \ref{tab:parameters}. The residuals show stability over the entire time span. Right panels: The same residuals as a function of orbital phase measured from the ascending node. There is no evident systematic variation of the residuals as a function of orbital phase due to delays in the companion atmospheres. This indicates that there are no eclipses, whose presence would be indicated by a clear excess delay around an orbital phase of 0.25.}
    \label{fig:residuals_3pulsars}
\end{figure*}

We obtained stable timing solutions for the three pulsars J0023+0923, J2214+3000 and J2234+0944 over time spans of 7$-$8 years, using data from four of the EPTA telescopes. The best-fit timing solutions for these three pulsars are shown in Table \ref{tab:parameters}, and in Fig. \ref{fig:residuals_3pulsars} the timing residuals for the entire time-span are shown.

\newpage
\begin{table*}
\caption{Timing solutions for PSRs~J0023+0923, J2214+3000 and J2234+0944. The uncertainty in the least significant quoted digit of a parameter is given in brackets. This is the nominal 1-$\sigma$ \texttt{TEMPO2} uncertainty. The reference clock scale, Solar System ephemeris and binary model used are: TT(TAI), DE421, and ELL1. The reduced $\chi^2$ for the solution is give before the \texttt{T2EFAC} was implemented. After the error-scaling was applied, the reduced $\chi^2\simeq 1$. It should be noted that the DM and $\dot{\mathrm{DM}}$ were only fitted to the low-frequency WSRT data, which are best suited for the determination of DM variations. The $P_b$ and $\dot{P_{b}}$ values for PSR J0023$-$0923 are derived from FB0 and FB1.}
\label{tab:parameters}
\begin{tabular}{lccc}
	\hline\hline
	\multicolumn{4}{c}{Fit and data-set} \\
	\hline
	Pulsar name\dotfill & J0023+0923 & J2214+3000 & J2234+0944 \\ 
	MJD range\dotfill & 55309.6---58167.8 & 55112.9---58168.8 & 55523.7---58160.5 \\ 
	Data span (yr)\dotfill & 7.83 & 8.37 & 7.22\\ 
	Number of TOAs\dotfill & 1073 & 648  & 986\\
	Weighted RMS timing residual ($\mu s$)\dotfill & 3.5 & 4.3 & 2.1\\ 
	Reduced $\chi^2$ value (before application of EFACs) \dotfill & 2.49 & 2.12 & 4.45\\
	\hline
	\multicolumn{4}{c}{Measured Quantities}  \\ 
	\hline
	Right ascension, $\alpha$ (hh:mm:ss)\dotfill & 00:23:16.877787(18) & 22:14:38.852794(14) & 22:34:46.853741(6) \\ 
	Declination, $\delta$ (dd:mm:ss)\dotfill & +09:23:23.8620(7) & +30:00:38.1964(3)  & +09:44:30.2564(2)\\ 
	Pulse frequency, $\nu$ (s$^{-1}$)\dotfill & 327.8470154891364(8)  & 320.5922873902387(20) & 275.7078283945944(6) \\ 
	First derivative of pulse frequency, $\dot{\nu}$ (s$^{-2}$)\dotfill & $-$1.22773(3)$\times 10^{-15}$ & $-$1.51368(5)$\times 10^{-15}$ & $-$1.527888(17)$\times 10^{-15}$\\ 
	Dispersion measure, DM (cm$^{-3}$pc)\dotfill & 14.32726(10) & 22.5519(6)  & 17.8253(7)\\ 
	First derivative of dispersion measure, $\dot{\mathrm{DM}}$ (cm$^{-3}$pc\,yr$^{-1}$)\dotfill & & 3.6(4)$\times 10^{-4}$ &  \\
	Proper motion in right ascension, $\mu_{\alpha} \cos \delta$ (mas\,yr$^{-1}$)\dotfill & $-$12.79(15) & 20.77(8) & 6.96(6) \\ 
	Proper motion in declination, $\mu_{\delta}$ (mas\,yr$^{-1}$)\dotfill &  $-$5.4(4)  & $-$1.46(12) &  $-$32.22(10) \\  
	Orbital period, $P_b$ (d)\dotfill &  & 0.4166329517(3) & 0.41966004343(7) \\ 
	Projected semi-major axis of orbit, $x$ (lt-s)\dotfill & 0.03484127(18) & 0.0590811(3) & 0.06842948(12)\\ 
	Orbital frequency, FB0 (s$^{-1}$)\dotfill & 8.33872167(8)$\times 10^{-5}$ & & \\ 
	First derivative of orbital frequency, FB1\dotfill & $-$3(2)$\times 10^{-20}$ & & \\ 
	Second derivative of orbital frequency, FB2\dotfill & $-$4(30)$\times 10^{-29}$ & & \\ 
	Third derivative of orbital frequency, FB3\dotfill & 3(2)$\times 10^{-36}$ \\
	Time of Ascending node, TASC (MJD)\dotfill & 55186.113622(4) & 55094.1380407(6) &  55517.4822998(3) \\ 
	EPS1\dotfill &   2.1(11)$\times 10^{-5}$  & 1.1(9)$\times 10^{-5}$ & 5.8(33)$\times 10^{-6}$ \\ 
	EPS2\dotfill & $-$5(100)$\times 10^{-7}$ & 1(1)$\times 10^{-5}$  & 8(3)$\times 10^{-6}$\\
	\hline
	\multicolumn{4}{c}{Upper Limits}  \\ 
	\hline
	First derivative of $x$, $\dot{x}$ ($10^{-12}$)\dotfill &  & $-$8.4(58)$\times 10^{-15}$ & \\
	\hline
	\multicolumn{4}{c}{Derived Quantities}  \\ 
	\hline
	Orbital period, $P_b$ (d)\dotfill & 0.138799141(1) &   &  \\
	First derivative of orbital period, $\dot{P_b}$\dotfill &  3.9(24)$\times 10^{-12}$  & &  \\
	Eccentricity ($e$)\dotfill & $2.1(25)\times 10^{-5}$ & $1.5(9)\times 10^{-5}$ & $9.6(33)\times 10^{-6}$\\
	Mass function, f (10$^{-6}$M$_\odot$)\dotfill & 2.35717(4) &  1.27561(2) & 1.95351(1) \\
	DM-derived distance (YMW16 model), d (kpc)\dotfill & 1.11 & 0.60 & 0.77 \\
	DM-derived distance (NE2001 model), d (kpc)\dotfill& 0.69 & 1.54 & 1.00 \\
	\hline
	\multicolumn{4}{c}{Set Quantities}  \\ 
	\hline
	Epoch of frequency and position determination (MJD)\dotfill & 56738 & 56640 & 56842\\ 
	Epoch of dispersion measure determination (MJD)\dotfill & 56452 & 56203 & 56763\\ 
	\hline
	\hline
\end{tabular}
\end{table*}
\newpage

\subsection{New parameter measurements}\label{sec:result:proper motion}

For low eccentricities the location of periastron is not well measured in the TOAs and a strong covariance between the longitude of periastron ($\omega$) and the epoch of periastron passage (T$_0$) occurs. This leads to very large uncertainties in the estimate of $\chi^2$ of these parameters \citep[see][]{Lange_2001MNRAS.326..274L}. This is avoided by using the ELL1 model in \texttt{tempo2}. 

When using the ELL1 model, the Laplace parameters, EPS1=$e\thinspace sin(\omega)$ and EPS2=$e\thinspace cos(\omega)$, instead of the eccentricity, $e$, and $\omega$, are included \citep{Hobbs_2006MNRAS.369..655H}. Furthermore T$_{\mathrm{ASC}}$ is included in the ELL1 model rather than T$_0$.
We measure EPS1 and EPS2, which are orthogonal components of the eccentricity. The eccentricity is given by $e$=$\sqrt{\mathrm{EPS1}^2+\mathrm{EPS2}^2}$ and the eccentricities of the pulsars are thus nominally $e$=2.1(25)$\times$10$^{-5}$, $e$=1.5(9)$\times$10$^{-5}$, $e$=9.6(33)$\times$10$^{-6}$ respectively for PSRs J0023+0923, J2214+3000 and J2234+0944.
Given the low significance, the eccentricities are interpreted as upper limits.

In this paper we also introduce and measure a first-order derivative of the dispersion measure ($\mathrm{\dot{DM}}=d\mathrm{DM}/dt$) for PSR J2214+3000, of 3.6(4)$\times$10$^{-4}$cm$^{-3}$\thinspace pc \thinspace yr$^{-1}$. This accounts for the DM variations we show in  Fig. \ref{fig:DM_plot} (for more details see Section \ref{sec:result:DM}). We further find a very marginal detection of the first-order derivative of the semi-major axis, of $-$8.4(58)$\times 10^{-15}$ for PSR J2214+3000. The semi-major axis marginal upper limit measurement of PSR J2214+3000 is not used in the final timing solution presented in Table \ref{tab:parameters}.
In addition we derived a significant value of $\dot{P_\mathrm{b}}$ for PSR J0023+0923, of 4.7(23)$\times 10^{-12}$, from the measured orbital frecuency (FB0) and first order orbital frequency derivative (FB1). $\dot{P_\mathrm{b}}$ is slightly larger than previous measurement \citep{Arzoumanian_2018ApJS..235...37A}, however, this measurement is still within the error on the parameter found in this paper. In general, our results are consistent to the results found in \citet{Arzoumanian_2018ApJS..235...37A}.

\subsection{Dispersion Measure variations }\label{sec:result:DM}

\begin{figure}
\centering
\includegraphics[width=1.0\linewidth]{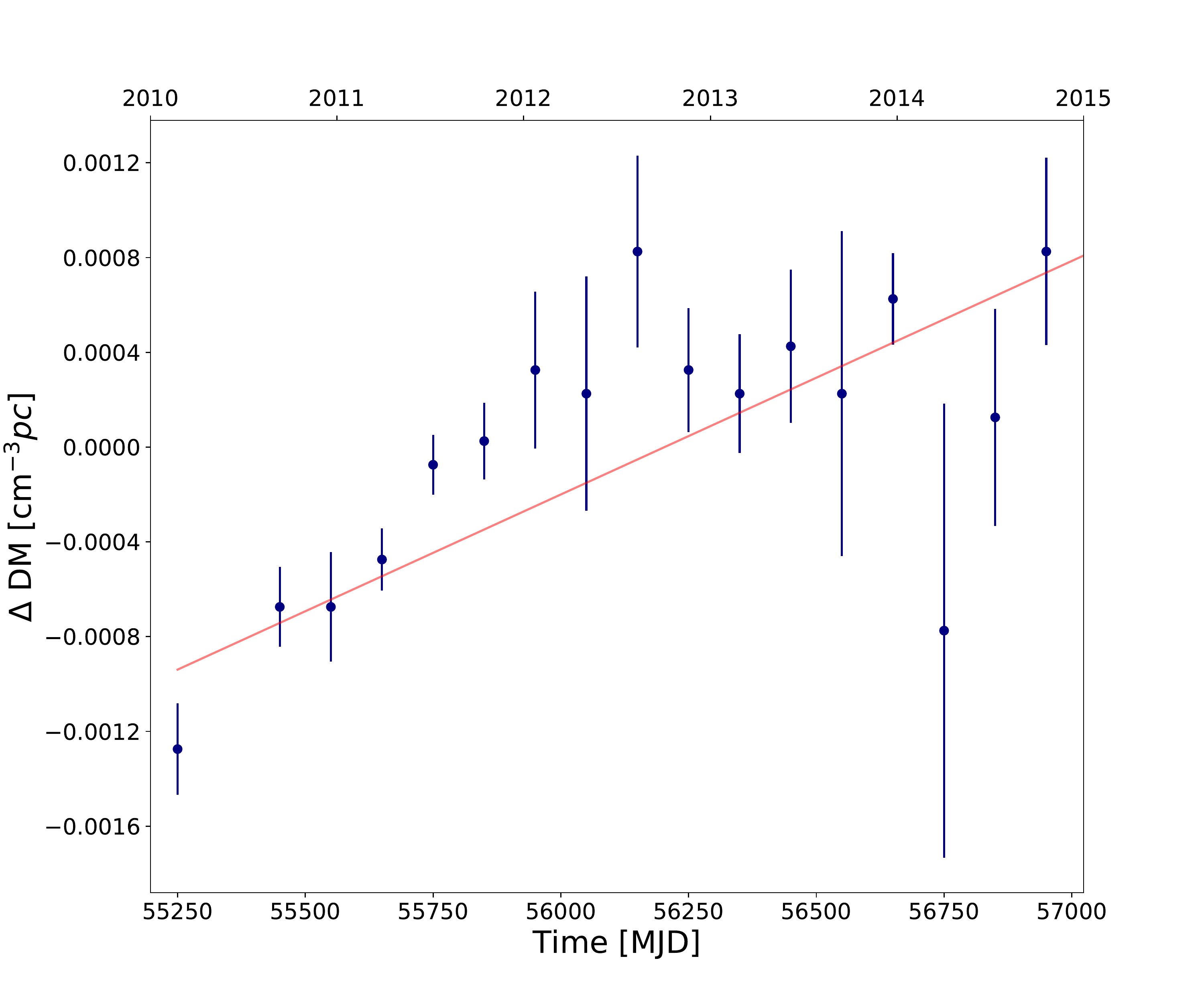}
\caption{Measured DM variations as a function of time, derived from the combined WSRT low- and high-frequency data of PSR~J2214+3000. A clear gradient in DM can be detected during approximately the first three years of our timing campaign. The red line indicates DM(t)=DM+(t-t$_0$)$\times \dot{\mathrm{DM}}$. $\dot{\mathrm{DM}}$ was measured only by a fit to the low frequency WSRT data. $\Delta$ DM was measured relative to the DM quoted in table \ref{tab:parameters} and was measured from fits of the DM to 100-200 day long segments.
}
\label{fig:DM_plot}
\end{figure}

The turbulent nature of the ionised ISM, which causes the scintillation described in Section \ref{sec:observations}, along with significant transverse velocities characteristic to many pulsars implies that the DM along the line of sight can vary significantly over a range of timescales. The effects of changes in the DM on the arrival times of the pulses is inversely proportional to the observing frequency squared, and are therefore more easily detectable at lower frequencies. DM variations are only detectable when the low-frequency data are included, as the DM uncertainty in the high-frequency data alone is too large to provide any sensitivity to DM variations. 
If we look at the combined low- and high-frequency WSRT data of PSR~J2214+3000 (see Fig. \ref{fig:DM_plot}), we detect DM variations. The DM measurements in Fig. \ref{fig:DM_plot} were obtained by fitting the DM to non-overlapping segments of data that were around 100-200 days long, to gain sufficient frequency coverage. 
In a separate analysis of the low-frequency  WSRT data, we detected a first-order DM derivative of 3.6(4)$\times$10$^{-4}$cm$^{-3}$\thinspace pc \thinspace yr$^{-1}$. For PSRs J0023+0923 and J2234+0944 we did not detect any DM variations. We furthermore tested whether the orbital phase was dependent on DM, see Section \ref{sec:disc:eclipses}.

\begin{figure}
\centering
\includegraphics[width=0.9\linewidth]{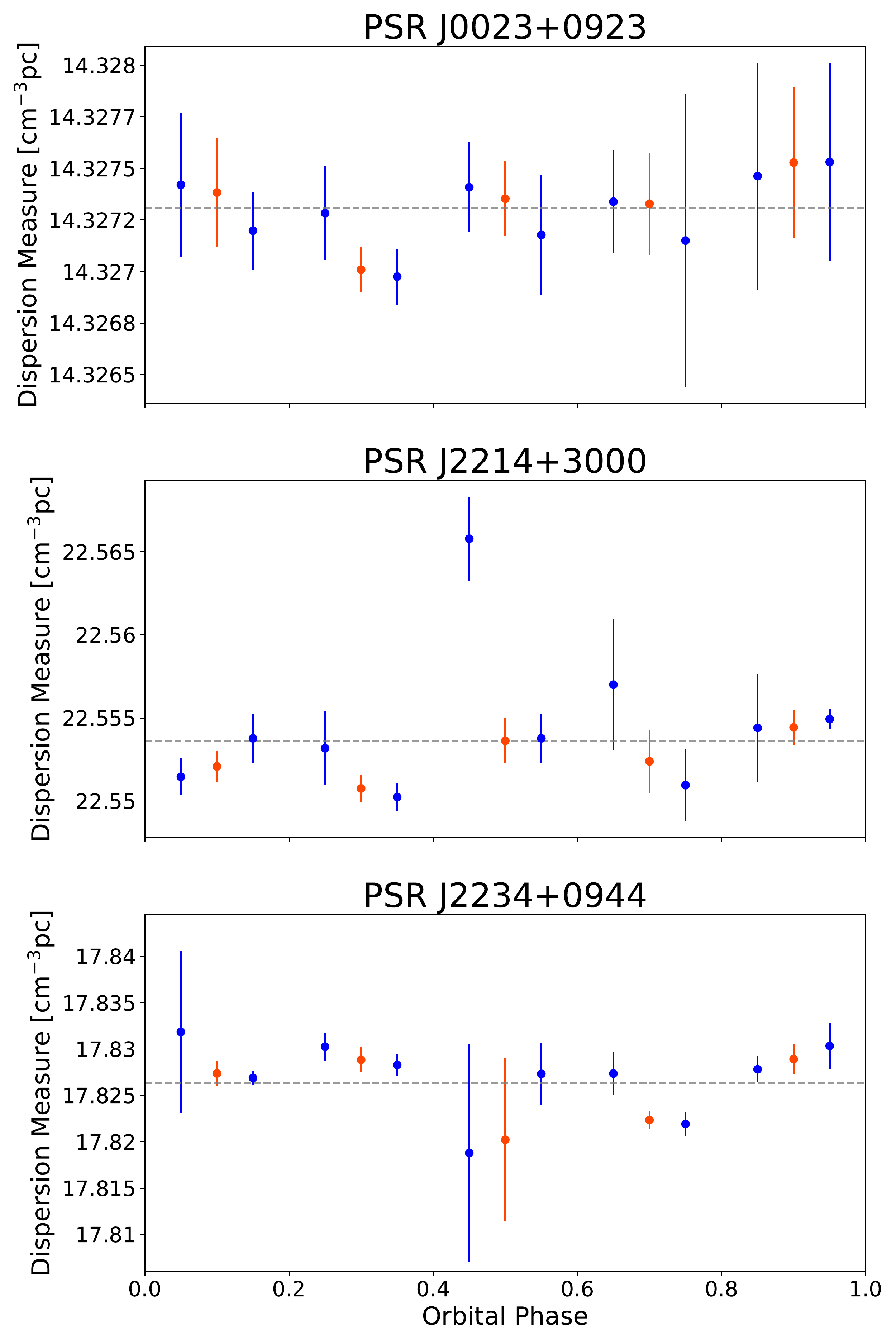}
\caption{Dispersion measure as a function of orbital phase. There appears to be no variation around the orbital phase of 0.25 where an eclipse would be expected.  
The plot shows measurements from the low frequency WSRT data. The blue points show the orbit sampled in steps of 0.1 P$_b$ while the red points are sampled in steps of 0.2 P$_b$. The blue points sample about 3-11 TOAs and the red points sample around 6-20 TOAs. The grey line denotes the average DM value.}
\label{fig:DM_plot_orbPhase}
\end{figure}

\section{Discussion}\label{sec:discussion}

In the following section we will discuss some of the features that the three sources show and relate these to the long-term stability of the three BWPs. Furthermore we will compare the properties of stable and unstable BWPs.

\subsection{Orbital Stability}\label{sed:discussion:stability}

In this paper, we classify a BW pulsar as stable if it has a single phase-connected timing solution, with only one orbital period derivative of fixed sign (see also \citet{Freire_2017MNRAS.471..857F}). However `instability' counter to this definition would not disqualify a BW pulsar as a PTA source. A more applicable definition for that case is provided in \citet{Bochenek_2015ApJ...813L...4B} and we do not elaborate on that any further in this article. We clarify that the instability as defined here is only linked to variability in the orbital period and not a red-noise process driven instability.

By this definition, the first BWP found, PSR~B1957+20, is unstable over long time spans, since it was not possible to create a timing solution spanning more than a few months to about a year \citep{Arzoumanian_1994ApJ...426L..85A}. The pulsar also shows radio eclipses due to the bloated companion and material surrounding the companion. The second BWP found, PSR J2051$-$0827, shows similar behaviour, including timing solutions that were only stable for short periods of time, of about 3\thinspace yrs \citep{Lazaridis_2011MNRAS.414.3134L, Shaifullah_2016MNRAS.462.1029S}. Contrary to the first two BWPs discovered, there is one BWP system PSR~J0610$-$2100, which shows a stable timing solution and is included in the EPTA and IPTA \citep{Desvignes_2016MNRAS.458.3341D, Verbiest_2016MNRAS.458.1267V}. 
 
The three BWPs analysed in this paper are very different from those early discoveries. None of these three BWPs show radio eclipses and it is possible to find a coherent timing solution for each of the three systems that is valid at least over a 7$-$8 year period. However, PSR J0023+0923 does show significant orbital period derivatives (see table \ref{tab:parameters}), and all orbital frequency (FB) parameters are non-zero, which seems to indicate that there are orbital variations present similar to what is found for e.g. PSR J2051$-$0827 \citep{Shaifullah_2016MNRAS.462.1029S}. However, all three pulsars have reasonably low timing residual rms ($\sim 2.1-4.3\mu s$) and all are good PTA sources.

\subsection{Eclipses}\label{sec:disc:eclipses}
We examined our data sets for the three pulsars for eclipses and did not find any evidence for radio eclipses for any of the three pulsars at any of our observed frequencies. See the right hand panels on Fig. \ref{fig:residuals_3pulsars}, which presents averages over many orbits. These plots indicate that there is not enough material around the companion to create the eclipses, as is also suggested for PSR J0023+0923 from X-ray observations by \citet{Gentile_2014ApJ...783...69G}. The lack of radio eclipses can be an indication that the companion stars are not filling their Roche lobes (see Section \ref{sec:disc:RL}).
To further test if we see any indications of eclipses, we studied the DM as a function of orbital period. 
If there was any indication of additional ionized material along the line of sight at around orbital phase 0.25 (i.e. when the companion is at the minimum distance from us and the MSPs is on the opposite part of the orbit), then it would be evident when looking at the DM as a function of the orbital phase \citep{Stappers_1996ApJ...465L.119S}. In Fig 5, we do not see any significant DM variation around orbital phase 0.25; however, we note that the orbital period is poorly sampled for all three pulsars in this paper. The poorly sampled orbital periods are the cause of the outliers in Fig. \ref{fig:DM_plot_orbPhase}.
Other evidence for eclipses not being present is seen in the mass function of the three black widows in this paper (see Table \ref{tab:parameters}). As mentioned by both \citet{Freire_2005ASPC..328..405F} and \citet{Guillemot_2019A&A...629A..92G} there seems to be a correlation between eclipses, or the lack thereof, and the mass function of the pulsars. The higher values of mass functions in general seem to belong to pulsars showing eclipses and the low mass functions, as we also see for the three pulsars in this paper, are often coinciding with BWs without eclipses (see table 2 in \citet{Guillemot_2019A&A...629A..92G}). 

\subsection{Roche-lobe filling factors}\label{sec:disc:RL}

The companion stars for the first-known two BWPs, PSRs B1957+20 and J2051$-$0827, are suggested to be Roche-lobe filling, with a Roche-lobe filling factor (RLFF) of about 0.9 (the filling factors are determined from fitting the optical light curve of the companion stars) \citep{Stappers_2001ApJ...548L.183S,Reynolds_2007MNRAS.379.1117R,Kerkwijk_2011ApJ...728...95V}. This means that the outer layers of the companion star are less gravitationally bound to the star and it is thus easier to expel material from, or ablate, the companion star. This is one of the essential details of the geometry of the BWPs, as it is the ablated material that causes the radio eclipses \citep{Kerkwijk_2011ApJ...728...95V}. 
Furthermore the BWP J1810+1744 similarly has a RLFF of about 0.8. This system also shows radio eclipses and instabilities \citep{Breton_2013ApJ...769..108B, Polzin_2018MNRAS.476.1968P}.\\

The BWP J2256$-$1024 was found by \cite{Breton_2013ApJ...769..108B} to have a relatively low RLFF of $\sim$0.4, but was reported by \cite{Gentile_2014ApJ...783...69G} to show orbital variability. Recent studies of another BWP, PSR J0636+5128 \citep{Kaplan_2018ApJ...864...15K}, show that this pulsar has a large RLFF, of about 0.75, but it does not show eclipses. The missing eclipses are probably due to the relatively low inclination angle of $\sim$24$^\circ$. This system seems to be a good candidate for orbital variations (Bak Nielsen in prep.). \\

Of the three pulsars discussed in this paper, the RLFF has only been measured for the companion star of PSR J0023+0923, and it is $\sim$0.3 \citep{Breton_2013ApJ...769..108B}. 
If the Roche lobe of a BWP system is not filled, eclipses would not be expected, which coincides with what is observed for PSRs~J0023+0924, J2214+3000 and J2234+0944. However, we do find that orbital variations are present for PSR J0023+0923. Thus it is evident that while RLFFs seem to be closely related to the inherent stability of a binary system, RLFF alone is not a good discriminator of timing solution stability.

\subsection{Additional intrinsic or observational parameters at play}
BWPs are still a rather heterogeneous group of objects and various factors impact on their observational properties as well as on the intrinsic process at play in these systems. Alongside the RLFF, the irradiation of the companion star by the pulsar and the distance between the pulsar and companion star are factors that have the potential to affect the orbital stability. For example the irradiation affects how heated and ablated the companion star is and thus how much it will potentially fill its Roche lobe \citep{Stappers_2001ApJ...548L.183S, Kerkwijk_2011ApJ...728...95V, Breton_2013ApJ...769..108B}. The distance to the BWPs affects the measured value of the RLFF, because it is determined by lightcurve modelling and is thus dependent on the apparent luminosity, and could thus affect our discussion in Section \ref{sec:disc:RL} \citep{Breton_2013ApJ...769..108B}. The orbital inclination of the system has no effect on the orbital stability, however, it does influence whether or not we see eclipses in these systems. The inclinations of PSRs J0023+0923, J2051$-$0827 and B1957+20 are similar, respectively $i\sim$58$^\circ$, $i\sim$40$^\circ$ and $i\sim$65$^\circ$, inferred from light curve modelling. The first one of these systems does not show eclipses and the other two do \citep{Breton_2013ApJ...769..108B,Stappers_2001ApJ...548L.183S,Kerkwijk_2011ApJ...728...95V}. However, one system where it is \textbf{likely} that the inclination angle does matter is J0636+5128, which has an inclination of $\sim$24$^\circ$, and shows no eclipses, even though it has a similar RLFF as PSRs J2051$-$0827 and B1957+20 \citep{Kaplan_2018ApJ...864...15K}.

\section{PSR J2234+0944 as a possible transitional system}\label{sec:disc:transitionalPSR}

Transitional millisecond pulsars are a relatively new class of binary pulsars, that show changes between emitting in radio and X-rays, and are believed to switch states between having an active accretion disc and no accretion disc \citep{Archibald_2009Sci...324.1411A}. Since the first transitional millisecond pulsar, PSR~J1023+0038, was discovered to switch states, two more have been found, and all three systems are RBPs \citep{Papitto_2013Natur.501..517P, Bassa_2014MNRAS.441.1825B, Roy_2015ApJ...800L..12R}. 
The companion stars of BWPs and RBPs are very different, as RBPs are thought to have non-degenerate companions with a higher mass than the semi-degenerate companions of BWPs \citep{Chen_2013ApJ...775...27C}. However, variations in their orbit and the presence of radio eclipses in some of these systems are characteristics shared between the BWPs and the transitional systems \citep{Fruchter_1988Natur.333..237F, Stappers_1996ApJ...473L.119S,Archibald_2009Sci...324.1411A, Papitto_2013Natur.501..517P, Chen_2013ApJ...775...27C,Roy_2015ApJ...800L..12R}.
\citet{Torres_2017ApJ...836...68T} suggested that PSR J2234+0944 could be a potential transitional system, based on variations in the $\gamma$-ray spectrum, which is similar to what is seen in the transitional millisecond pulsar J1227$-$4853. However, PSR J2234+0944 has a timing solution that is stable over 8\thinspace yrs, with no obvious irregularities or eclipses. Given the data, this system appears to be different from the variable transitional systems identified so far. 

\section{Conclusions}\label{sec:conclusion}

We studied three BWPs: PSRs~J0023+0923, J2234+0944, and J2214+3000, all of which show different behaviour than the first two BWPs discovered, PSRs~B1957+20 and J2051$-$0827, that are counted as unstable systems. 
We present timing solutions for PSRs~J0023+0923, J2214+3000 and J2234+0944 over a data span of 7$-$8\thinspace yrs. PSR J0023+0923 shows orbital variations, which can be modelled with up to a third order orbital frequency derivative (see Table \ref{tab:parameters}). That seems to counter the argument that a low value of the Roche Lobe Filling Factor is an indicator for stable BWPs. From the orbital frequency derivatives we derived an orbital period derivative for PSR J0023+0923, of 3.9(24)$\times 10^{-12}$
The other two BWPs in this study, PSRs J2214+3000 and J2234+0944, show no orbital variations. It is possible that this is due to their companion star not filling its Roche lobe. This would make the companion stars more gravitationally tightly bound to themself, resulting in a cleaner system with less material to interfere with, or to affect the radio signal and the orbital parameters.
Since the three pulsars show stability, or only minimal instability, over long timescales it is possible to use the systems discussed here as part of a pulsar timing array, and all three pulsars have been added to the EPTA source list.
It remains to be seen if the lack of orbital stability in some of these gravitationally tightly bound systems is tied to specific high-energy signatures. 
Comparing with more BWPs and other pulsar systems will provide a clearer picture of what the most important factors are for determining why a BWP system has a stable or unstable orbital solution. 

\section*{Acknowledgements}

The European Pulsar Timing Array (EPTA: www.epta.eu.org) is a collaboration of European institutes to work towards the direct detection of low-frequency gravitational waves and the running of the Large European Array for Pulsars (LEAP).
The Westerbork Synthesis Radio Telescope is operated by the Netherlands Institute for Radio Astronomy (ASTRON) with support from The Netherlands Foundation for Scientific Research NWO. Pulsar research at the Jodrell Bank Centre for Astrophysics and the observations using the Lovell Telescope is supported by a consolidated grant from the STFC in the UK. The Nan\c{c}ay Radio Observatory is operated by the Paris Observatory, associated with the French Centre National de la Recherche Scientifique (CNRS) and with the Universit\'{e} d'Orl\'{e}ans. We acknowledge financial support from the `Gravitation, R\'{e}f\'{e}rences, Astronomie, M\'{e}trologie' (GRAM) national programme of CNRS/INSU, France. Part of this work is based on observations with the 100-m telescope of the Max-Planck-Institut f\"{u}r Radioastronomie (MPIfR) at Effelsberg. 
GJ and GS acknowledge support from the Netherlands Organisation for Scientific Research NWO (TOP2.614.001.602).
GD gratefully acknowledges support from European Research Council (ERC) Synergy Grant ``BlackHoleCam" Grant Agreement Number 610058.
The authors also thank Stefan Os\l{}owski for contributions to observations used in this paper.




\bibliographystyle{mnras}
\bibliography{bibliography} 

\begin{thebibliography}{}
\makeatletter
\relax
\def\mn@urlcharsother{\let\do\@makeother \do\$\do\&\do\#\do\^\do\_\do\%\do\~}
\def\mn@doi{\begingroup\mn@urlcharsother \@ifnextchar [ {\mn@doi@}
  {\mn@doi@[]}}
\def\mn@doi@[#1]#2{\def\@tempa{#1}\ifx\@tempa\@empty \href
  {http://dx.doi.org/#2} {doi:#2}\else \href {http://dx.doi.org/#2} {#1}\fi
  \endgroup}
\def\mn@eprint#1#2{\mn@eprint@#1:#2::\@nil}
\def\mn@eprint@arXiv#1{\href {http://arxiv.org/abs/#1} {{\tt arXiv:#1}}}
\def\mn@eprint@dblp#1{\href {http://dblp.uni-trier.de/rec/bibtex/#1.xml}
  {dblp:#1}}
\def\mn@eprint@#1:#2:#3:#4\@nil{\def\@tempa {#1}\def\@tempb {#2}\def\@tempc
  {#3}\ifx \@tempc \@empty \let \@tempc \@tempb \let \@tempb \@tempa \fi \ifx
  \@tempb \@empty \def\@tempb {arXiv}\fi \@ifundefined
  {mn@eprint@\@tempb}{\@tempb:\@tempc}{\expandafter \expandafter \csname
  mn@eprint@\@tempb\endcsname \expandafter{\@tempc}}}

\bibitem[\protect\citeauthoryear{{Abdo} et~al.,}{{Abdo}
  et~al.}{2013}]{Abdo_2013ApJS..208...17A}
{Abdo} A.~A.,  et~al., 2013, \mn@doi [\apjs] {10.1088/0067-0049/208/2/17},
  \href {https://ui.adsabs.harvard.edu/abs/2013ApJS..208...17A} {208, 17}

\bibitem[\protect\citeauthoryear{{Al Noori} et~al.,}{{Al Noori}
  et~al.}{2018}]{AlNoori_2018ApJ...861...89A}
{Al Noori} H.,  et~al., 2018, \mn@doi [\apj] {10.3847/1538-4357/aac828}, \href
  {https://ui.adsabs.harvard.edu/abs/2018ApJ...861...89A} {861, 89}

\bibitem[\protect\citeauthoryear{{Alpar}, {Cheng}, {Ruderman}  \&
  {Shaham}}{{Alpar} et~al.}{1982}]{Alpar_1982Natur.300..728A}
{Alpar} M.~A.,  {Cheng} A.~F.,  {Ruderman} M.~A.,   {Shaham} J.,  1982, \mn@doi
  [\nat] {10.1038/300728a0}, \href
  {http://adsabs.harvard.edu/abs/1982Natur.300..728A} {300, 728}

\bibitem[\protect\citeauthoryear{{Archibald} et~al.,}{{Archibald}
  et~al.}{2009}]{Archibald_2009Sci...324.1411A}
{Archibald} A.~M.,  et~al., 2009, \mn@doi [Science] {10.1126/science.1172740},
  \href {http://adsabs.harvard.edu/abs/2009Sci...324.1411A} {324, 1411}

\bibitem[\protect\citeauthoryear{{Arzoumanian}, {Fruchter}  \&
  {Taylor}}{{Arzoumanian} et~al.}{1994}]{Arzoumanian_1994ApJ...426L..85A}
{Arzoumanian} Z.,  {Fruchter} A.~S.,   {Taylor} J.~H.,  1994, \mn@doi [\apjl]
  {10.1086/187346}, \href {http://adsabs.harvard.edu/abs/1994ApJ...426L..85A}
  {426, 85}

\bibitem[\protect\citeauthoryear{{Arzoumanian} et~al.,}{{Arzoumanian}
  et~al.}{2018}]{Arzoumanian_2018ApJS..235...37A}
{Arzoumanian} Z.,  et~al., 2018, \mn@doi [\apjs] {10.3847/1538-4365/aab5b0},
  \href {https://ui.adsabs.harvard.edu/abs/2018ApJS..235...37A} {235, 37}

\bibitem[\protect\citeauthoryear{{Bassa} et~al.,}{{Bassa}
  et~al.}{2014}]{Bassa_2014MNRAS.441.1825B}
{Bassa} C.~G.,  et~al., 2014, \mn@doi [\mnras] {10.1093/mnras/stu708}, \href
  {http://adsabs.harvard.edu/abs/2014MNRAS.441.1825B} {441, 1825}

\bibitem[\protect\citeauthoryear{{Bassa} et~al.,}{{Bassa}
  et~al.}{2016}]{Bassa_2016MNRAS.456.2196B}
{Bassa} C.~G.,  et~al., 2016, \mn@doi [\mnras] {10.1093/mnras/stv2755}, \href
  {https://ui.adsabs.harvard.edu/abs/2016MNRAS.456.2196B} {456, 2196}

\bibitem[\protect\citeauthoryear{{Bhattacharya} \& {van den
  Heuvel}}{{Bhattacharya} \& {van den
  Heuvel}}{1991}]{Bhattacharya_1991PhR...203....1B}
{Bhattacharya} D.,  {van den Heuvel} E.~P.~J.,  1991, \mn@doi [\physrep]
  {10.1016/0370-1573(91)90064-S}, \href
  {http://adsabs.harvard.edu/abs/1991PhR...203....1B} {203, 1}

\bibitem[\protect\citeauthoryear{{Bochenek}, {Ransom}  \&
  {Demorest}}{{Bochenek} et~al.}{2015}]{Bochenek_2015ApJ...813L...4B}
{Bochenek} C.,  {Ransom} S.,   {Demorest} P.,  2015, \mn@doi [\apjl]
  {10.1088/2041-8205/813/1/L4}, \href
  {http://adsabs.harvard.edu/abs/2015ApJ...813L...4B} {813, L4}

\bibitem[\protect\citeauthoryear{{Breton} et~al.,}{{Breton}
  et~al.}{2013}]{Breton_2013ApJ...769..108B}
{Breton} R.~P.,  et~al., 2013, \mn@doi [\apj] {10.1088/0004-637X/769/2/108},
  \href {http://adsabs.harvard.edu/abs/2013ApJ...769..108B} {769, 108}

\bibitem[\protect\citeauthoryear{{Chen}, {Chen}, {Tauris}  \& {Han}}{{Chen}
  et~al.}{2013}]{Chen_2013ApJ...775...27C}
{Chen} H.-L.,  {Chen} X.,  {Tauris} T.~M.,   {Han} Z.,  2013, \mn@doi [\apj]
  {10.1088/0004-637X/775/1/27}, \href
  {http://adsabs.harvard.edu/abs/2013ApJ...775...27C} {775, 27}

\bibitem[\protect\citeauthoryear{{Desvignes}, {Barott}, {Cognard}, {Lespagnol}
  \& {Theureau}}{{Desvignes} et~al.}{2011}]{Desvignes_2011AIPC.1357..349D}
{Desvignes} G.,  {Barott} W.~C.,  {Cognard} I.,  {Lespagnol} P.,   {Theureau}
  G.,  2011, in {Burgay} M.,  {D'Amico} N.,  {Esposito} P.,  {Pellizzoni} A.,
  {Possenti} A.,  eds,  American Institute of Physics Conference Series Vol.
  1357, American Institute of Physics Conference Series. pp 349--350,
  \mn@doi{10.1063/1.3615154}

\bibitem[\protect\citeauthoryear{{Desvignes} et~al.,}{{Desvignes}
  et~al.}{2016}]{Desvignes_2016MNRAS.458.3341D}
{Desvignes} G.,  et~al., 2016, \mn@doi [\mnras] {10.1093/mnras/stw483}, \href
  {http://adsabs.harvard.edu/abs/2016MNRAS.458.3341D} {458, 3341}

\bibitem[\protect\citeauthoryear{{Ferraro}, {Possenti}, {D'Amico}  \&
  {Sabbi}}{{Ferraro} et~al.}{2001}]{Ferraro_2001ApJ...561L..93F}
{Ferraro} F.~R.,  {Possenti} A.,  {D'Amico} N.,   {Sabbi} E.,  2001, \mn@doi
  [\apjl] {10.1086/324563}, \href
  {https://ui.adsabs.harvard.edu/abs/2001ApJ...561L..93F} {561, L93}

\bibitem[\protect\citeauthoryear{{Folkner}, {Williams}  \& {Boggs}}{{Folkner}
  et~al.}{2009}]{Folkner_2009IPNPR.178C...1F}
{Folkner} W.~M.,  {Williams} J.~G.,   {Boggs} D.~H.,  2009, Interplanetary
  Network Progress Report, \href
  {http://adsabs.harvard.edu/abs/2009IPNPR.178C...1F} {178, 1}

\bibitem[\protect\citeauthoryear{{Foster} \& {Backer}}{{Foster} \&
  {Backer}}{1990}]{Foster_1990BAAS...22.1341F}
{Foster} R.~S.,  {Backer} D.~C.,  1990, in \baas. p.~1341

\bibitem[\protect\citeauthoryear{{Freire}}{{Freire}}{2005}]{Freire_2005ASPC..328..405F}
{Freire} P.~C.~C.,  2005, in {Rasio} F.~A.,  {Stairs} I.~H.,  eds,
  Astronomical Society of the Pacific Conference Series Vol. 328, Binary Radio
  Pulsars. p.~405 (\mn@eprint {arXiv} {astro-ph/0404105})

\bibitem[\protect\citeauthoryear{{Freire} et~al.,}{{Freire}
  et~al.}{2017}]{Freire_2017MNRAS.471..857F}
{Freire} P.~C.~C.,  et~al., 2017, \mn@doi [\mnras] {10.1093/mnras/stx1533},
  \href {https://ui.adsabs.harvard.edu/abs/2017MNRAS.471..857F} {471, 857}

\bibitem[\protect\citeauthoryear{{Fruchter}, {Stinebring}  \&
  {Taylor}}{{Fruchter} et~al.}{1988}]{Fruchter_1988Natur.333..237F}
{Fruchter} A.~S.,  {Stinebring} D.~R.,   {Taylor} J.~H.,  1988, \mn@doi [\nat]
  {10.1038/333237a0}, \href {http://adsabs.harvard.edu/abs/1988Natur.333..237F}
  {333, 237}

\bibitem[\protect\citeauthoryear{{Gentile} et~al.,}{{Gentile}
  et~al.}{2014}]{Gentile_2014ApJ...783...69G}
{Gentile} P.~A.,  et~al., 2014, \mn@doi [\apj] {10.1088/0004-637X/783/2/69},
  \href {http://adsabs.harvard.edu/abs/2014ApJ...783...69G} {783, 69}

\bibitem[\protect\citeauthoryear{{Guillemot} et~al.,}{{Guillemot}
  et~al.}{2016}]{Guillemot_2016A&A...587A.109G}
{Guillemot} L.,  et~al., 2016, \mn@doi [\aap] {10.1051/0004-6361/201527847},
  \href {http://adsabs.harvard.edu/abs/2016A%26A...587A.109G} {587, A109}

\bibitem[\protect\citeauthoryear{{Guillemot}, {Octau}, {Cognard}, {Desvignes},
  {Freire}, {Smith}, {Theureau}  \& {Burnett}}{{Guillemot}
  et~al.}{2019}]{Guillemot_2019A&A...629A..92G}
{Guillemot} L.,  {Octau} F.,  {Cognard} I.,  {Desvignes} G.,  {Freire}
  P.~C.~C.,  {Smith} D.~A.,  {Theureau} G.,   {Burnett} T.~H.,  2019, \mn@doi
  [\aap] {10.1051/0004-6361/201936015}, \href
  {https://ui.adsabs.harvard.edu/abs/2019A&A...629A..92G} {629, A92}

\bibitem[\protect\citeauthoryear{{Hessels} et~al.,}{{Hessels}
  et~al.}{2011}]{Hessels_2011AIPC.1357...40H}
{Hessels} J.~W.~T.,  et~al., 2011, in {Burgay} M.,  {D'Amico} N.,  {Esposito}
  P.,  {Pellizzoni} A.,   {Possenti} A.,  eds,  American Institute of Physics
  Conference Series Vol. 1357, American Institute of Physics Conference Series.
  pp 40--43 (\mn@eprint {arXiv} {1101.1742}), \mn@doi{10.1063/1.3615072}

\bibitem[\protect\citeauthoryear{{Hewish}, {Bell}, {Pilkington}, {Scott}  \&
  {Collins}}{{Hewish} et~al.}{1968}]{Hewish_1968Natur.217..709H}
{Hewish} A.,  {Bell} S.~J.,  {Pilkington} J.~D.~H.,  {Scott} P.~F.,   {Collins}
  R.~A.,  1968, \mn@doi [\nat] {10.1038/217709a0}, \href
  {http://adsabs.harvard.edu/abs/1968Natur.217..709H} {217, 709}

\bibitem[\protect\citeauthoryear{{Hobbs}, {Edwards}  \& {Manchester}}{{Hobbs}
  et~al.}{2006}]{Hobbs_2006MNRAS.369..655H}
{Hobbs} G.~B.,  {Edwards} R.~T.,   {Manchester} R.~N.,  2006, \mn@doi [\mnras]
  {10.1111/j.1365-2966.2006.10302.x}, \href
  {http://adsabs.harvard.edu/abs/2006MNRAS.369..655H} {369, 655}

\bibitem[\protect\citeauthoryear{{Kaplan}, {Stovall}, {van Kerkwijk},
  {Fremling}  \& {Istrate}}{{Kaplan} et~al.}{2018}]{Kaplan_2018ApJ...864...15K}
{Kaplan} D.~L.,  {Stovall} K.,  {van Kerkwijk} M.~H.,  {Fremling} C.,
  {Istrate} A.~G.,  2018, \mn@doi [\apj] {10.3847/1538-4357/aad54c}, \href
  {http://adsabs.harvard.edu/abs/2018ApJ...864...15K} {864, 15}

\bibitem[\protect\citeauthoryear{{Karuppusamy}, {Stappers}  \& {van
  Straten}}{{Karuppusamy} et~al.}{2008}]{Karuppusamy_2008PASP..120..191K}
{Karuppusamy} R.,  {Stappers} B.,   {van Straten} W.,  2008, \mn@doi [\pasp]
  {10.1086/528699}, \href {http://adsabs.harvard.edu/abs/2008PASP..120..191K}
  {120, 191}

\bibitem[\protect\citeauthoryear{{Lange}, {Camilo}, {Wex}, {Kramer}, {Backer},
  {Lyne}  \& {Doroshenko}}{{Lange} et~al.}{2001}]{Lange_2001MNRAS.326..274L}
{Lange} C.,  {Camilo} F.,  {Wex} N.,  {Kramer} M.,  {Backer} D.~C.,  {Lyne}
  A.~G.,   {Doroshenko} O.,  2001, \mn@doi [\mnras]
  {10.1046/j.1365-8711.2001.04606.x}, \href
  {http://adsabs.harvard.edu/abs/2001MNRAS.326..274L} {326, 274}

\bibitem[\protect\citeauthoryear{{Lazaridis} et~al.,}{{Lazaridis}
  et~al.}{2011}]{Lazaridis_2011MNRAS.414.3134L}
{Lazaridis} K.,  et~al., 2011, \mn@doi [\mnras]
  {10.1111/j.1365-2966.2011.18610.x}, \href
  {http://adsabs.harvard.edu/abs/2011MNRAS.414.3134L} {414, 3134}

\bibitem[\protect\citeauthoryear{{Lazarus}, {Karuppusamy}, {Graikou},
  {Caballero}, {Champion}, {Lee}, {Verbiest}  \& {Kramer}}{{Lazarus}
  et~al.}{2016}]{Lazarus_2016MNRAS.458..868L}
{Lazarus} P.,  {Karuppusamy} R.,  {Graikou} E.,  {Caballero} R.~N.,  {Champion}
  D.~J.,  {Lee} K.~J.,  {Verbiest} J.~P.~W.,   {Kramer} M.,  2016, \mn@doi
  [\mnras] {10.1093/mnras/stw189}, \href
  {http://adsabs.harvard.edu/abs/2016MNRAS.458..868L} {458, 868}

\bibitem[\protect\citeauthoryear{{Manchester}, {Hobbs}, {Teoh}  \&
  {Hobbs}}{{Manchester} et~al.}{2005}]{Manchester_2005yCat.7245}
{Manchester} R.~N.,  {Hobbs} G.~B.,  {Teoh} A.,   {Hobbs} M.,  2005, VizieR
  Online Data Catalog, \href
  {https://ui.adsabs.harvard.edu/abs/2005yCat.7245....0M} {p. VII/245}

\bibitem[\protect\citeauthoryear{{Nita} \& {Gary}}{{Nita} \&
  {Gary}}{2010}]{Nita_2010MNRAS.406L..60N}
{Nita} G.~M.,  {Gary} D.~E.,  2010, \mn@doi [\mnras]
  {10.1111/j.1745-3933.2010.00882.x}, \href
  {https://ui.adsabs.harvard.edu/abs/2010MNRAS.406L..60N} {406, L60}

\bibitem[\protect\citeauthoryear{{Papitto} et~al.,}{{Papitto}
  et~al.}{2013}]{Papitto_2013Natur.501..517P}
{Papitto} A.,  et~al., 2013, \mn@doi [\nat] {10.1038/nature12470}, \href
  {http://adsabs.harvard.edu/abs/2013Natur.501..517P} {501, 517}

\bibitem[\protect\citeauthoryear{{Polzin} et~al.,}{{Polzin}
  et~al.}{2018}]{Polzin_2018MNRAS.476.1968P}
{Polzin} E.~J.,  et~al., 2018, \mn@doi [\mnras] {10.1093/mnras/sty349}, \href
  {http://adsabs.harvard.edu/abs/2018MNRAS.476.1968P} {476, 1968}

\bibitem[\protect\citeauthoryear{{Ransom} et~al.,}{{Ransom}
  et~al.}{2011}]{Ransom_2011ApJ...727L..16R}
{Ransom} S.~M.,  et~al., 2011, \mn@doi [\apjl] {10.1088/2041-8205/727/1/L16},
  \href {http://adsabs.harvard.edu/abs/2011ApJ...727L..16R} {727, L16}

\bibitem[\protect\citeauthoryear{{Ray} et~al.,}{{Ray}
  et~al.}{2012}]{Ray_2012arXiv1205.3089R}
{Ray} P.~S.,  et~al., 2012, preprint, \href
  {http://adsabs.harvard.edu/abs/2012arXiv1205.3089R} {} (\mn@eprint {arXiv}
  {1205.3089})

\bibitem[\protect\citeauthoryear{{Reynolds}, {Callanan}, {Fruchter}, {Torres},
  {Beer}  \& {Gibbons}}{{Reynolds} et~al.}{2007}]{Reynolds_2007MNRAS.379.1117R}
{Reynolds} M.~T.,  {Callanan} P.~J.,  {Fruchter} A.~S.,  {Torres} M.~A.~P.,
  {Beer} M.~E.,   {Gibbons} R.~A.,  2007, \mn@doi [\mnras]
  {10.1111/j.1365-2966.2007.11991.x}, \href
  {http://adsabs.harvard.edu/abs/2007MNRAS.379.1117R} {379, 1117}

\bibitem[\protect\citeauthoryear{{Rickett}}{{Rickett}}{1969}]{Rickett_1969Natur.221..158R}
{Rickett} B.~J.,  1969, \mn@doi [\nat] {10.1038/221158a0}, \href
  {https://ui.adsabs.harvard.edu/abs/1969Natur.221..158R} {221, 158}

\bibitem[\protect\citeauthoryear{{Roberts}}{{Roberts}}{2013}]{Roberts_2013IAUS..291..127R}
{Roberts} M.~S.~E.,  2013, in {van Leeuwen} J.,  ed.,  IAU Symposium Vol. 291,
  Neutron Stars and Pulsars: Challenges and Opportunities after 80 years. pp
  127--132 (\mn@eprint {arXiv} {1210.6903}), \mn@doi{10.1017/S174392131202337X}

\bibitem[\protect\citeauthoryear{{Roy} et~al.,}{{Roy}
  et~al.}{2015}]{Roy_2015ApJ...800L..12R}
{Roy} J.,  et~al., 2015, \mn@doi [\apjl] {10.1088/2041-8205/800/1/L12}, \href
  {http://adsabs.harvard.edu/abs/2015ApJ...800L..12R} {800, L12}

\bibitem[\protect\citeauthoryear{{Schroeder} \& {Halpern}}{{Schroeder} \&
  {Halpern}}{2014}]{Schroeder_2014ApJ...793...78S}
{Schroeder} J.,  {Halpern} J.,  2014, \mn@doi [\apj]
  {10.1088/0004-637X/793/2/78}, \href
  {http://adsabs.harvard.edu/abs/2014ApJ...793...78S} {793, 78}

\bibitem[\protect\citeauthoryear{{Shaifullah} et~al.,}{{Shaifullah}
  et~al.}{2016}]{Shaifullah_2016MNRAS.462.1029S}
{Shaifullah} G.,  et~al., 2016, \mn@doi [\mnras] {10.1093/mnras/stw1737}, \href
  {http://adsabs.harvard.edu/abs/2016MNRAS.462.1029S} {462, 1029}

\bibitem[\protect\citeauthoryear{{Stappers} et~al.,}{{Stappers}
  et~al.}{1996a}]{Stappers_1996ApJ...465L.119S}
{Stappers} B.~W.,  et~al., 1996a, \mn@doi [\apjl] {10.1086/310148}, \href
  {https://ui.adsabs.harvard.edu/abs/1996ApJ...465L.119S} {465, L119}

\bibitem[\protect\citeauthoryear{{Stappers}, {Bessell}  \& {Bailes}}{{Stappers}
  et~al.}{1996b}]{Stappers_1996ApJ...473L.119S}
{Stappers} B.~W.,  {Bessell} M.~S.,   {Bailes} M.,  1996b, \mn@doi [\apjl]
  {10.1086/310397}, \href {http://adsabs.harvard.edu/abs/1996ApJ...473L.119S}
  {473, L119}

\bibitem[\protect\citeauthoryear{{Stappers}, {Bailes}, {Manchester}, {Sandhu}
  \& {Toscano}}{{Stappers} et~al.}{1998}]{Stappers_1998ApJ...499L.183S}
{Stappers} B.~W.,  {Bailes} M.,  {Manchester} R.~N.,  {Sandhu} J.~S.,
  {Toscano} M.,  1998, \mn@doi [\apjl] {10.1086/311382}, \href
  {http://adsabs.harvard.edu/abs/1998ApJ...499L.183S} {499, L183}

\bibitem[\protect\citeauthoryear{{Stappers}, {van Kerkwijk}, {Bell}  \&
  {Kulkarni}}{{Stappers} et~al.}{2001}]{Stappers_2001ApJ...548L.183S}
{Stappers} B.~W.,  {van Kerkwijk} M.~H.,  {Bell} J.~F.,   {Kulkarni} S.~R.,
  2001, \mn@doi [\apjl] {10.1086/319106}, \href
  {http://adsabs.harvard.edu/abs/2001ApJ...548L.183S} {548, L183}

\bibitem[\protect\citeauthoryear{{Taylor}}{{Taylor}}{1992}]{Taylor_1992RSPTA.341..117T}
{Taylor} J.~H.,  1992, \mn@doi [Philosophical Transactions of the Royal Society
  of London Series A] {10.1098/rsta.1992.0088}, \href
  {https://ui.adsabs.harvard.edu/abs/1992RSPTA.341..117T} {341, 117}

\bibitem[\protect\citeauthoryear{{Tiburzi}}{{Tiburzi}}{2018}]{Tiburzi_2018PASA...35...13T}
{Tiburzi} C.,  2018, \mn@doi [\pasa] {10.1017/pasa.2018.7}, \href
  {https://ui.adsabs.harvard.edu/abs/2018PASA...35...13T} {35, e013}

\bibitem[\protect\citeauthoryear{{Torres}, {Ji}, {Li}, {Papitto}, {Rea}, {de
  O{\~n}a Wilhelmi}  \& {Zhang}}{{Torres}
  et~al.}{2017}]{Torres_2017ApJ...836...68T}
{Torres} D.~F.,  {Ji} L.,  {Li} J.,  {Papitto} A.,  {Rea} N.,  {de O{\~n}a
  Wilhelmi} E.,   {Zhang} S.,  2017, \mn@doi [\apj]
  {10.3847/1538-4357/836/1/68}, \href
  {http://adsabs.harvard.edu/abs/2017ApJ...836...68T} {836, 68}

\bibitem[\protect\citeauthoryear{{Verbiest} et~al.,}{{Verbiest}
  et~al.}{2016}]{Verbiest_2016MNRAS.458.1267V}
{Verbiest} J.~P.~W.,  et~al., 2016, \mn@doi [\mnras] {10.1093/mnras/stw347},
  \href {http://adsabs.harvard.edu/abs/2016MNRAS.458.1267V} {458, 1267}

\bibitem[\protect\citeauthoryear{{van Kerkwijk}, {Breton}  \& {Kulkarni}}{{van
  Kerkwijk} et~al.}{2011}]{Kerkwijk_2011ApJ...728...95V}
{van Kerkwijk} M.~H.,  {Breton} R.~P.,   {Kulkarni} S.~R.,  2011, \mn@doi
  [\apj] {10.1088/0004-637X/728/2/95}, \href
  {http://adsabs.harvard.edu/abs/2011ApJ...728...95V} {728, 95}

\bibitem[\protect\citeauthoryear{{van Straten}, {Demorest}  \& {Oslowski}}{{van
  Straten} et~al.}{2012}]{vanStraten_2012AR&T....9..237V}
{van Straten} W.,  {Demorest} P.,   {Oslowski} S.,  2012, Astronomical Research
  and Technology, \href {http://adsabs.harvard.edu/abs/2012AR%26T....9..237V}
  {9, 237}

\makeatother
\end{thebibliography}


\bsp	
\label{lastpage}
\end{document}